\documentstyle[12pt]{article}
\def\be{\begin{equation}}
\def\ee{\end{equation}}
\def\ba{\begin{eqnarray}}
\def\ea{\end{eqnarray}}

\def\fun#1#2{\lower3.6pt\vbox{\baselineskip0pt\lineskip.9pt
\ialign{$\mathsurround=0pt#1\hfill##\hfil$\crcr#2\crcr\sim\crcr}}}

\begin{document}

\begin{titlepage}
\null\vspace{-62pt}
\begin{flushright} UMD-PP-96-17\\
 hep-ph/9508365 \\ August 1995 \\
\end{flushright}
\vspace{0.2in}

\centerline{{\large \bf  $CP$ violation in multi-Higgs }}
\vspace{0.05in}
\centerline{{\large \bf  supersymmetric models } }

\vspace{0.2in}
\centerline{
Manuel Masip$^{(a,b)}$ \ \ and \ \ Andrija Ra\v{s}in$^{(a)}$ }
\vspace{0.2in}
\centerline{\it $^{(a)}$Department of Physics}
\centerline{\it University of Maryland}
\centerline{\it College Park, MD 20742, U.S.A.}
\vspace{.1in}
\centerline{\it $^{(b)}$Departamento de F\'\i sica
Te\'{o}rica y del Cosmos}
\centerline{\it Universidad de Granada}
\centerline{\it 18071 Granada, Spain}
\vspace{.3in}
\baselineskip=17pt

\centerline{\bf Abstract}
\begin{quotation}

We consider supersymmetric extensions of the standard model with two pairs
of Higgs doublets. We study the possibility of spontaneous $CP$ violation
in these scenarios and present a model where the origin of $CP$ violation
is soft, with all the complex phases in the Lagrangian derived from complex
masses and vacuum expectation values (VEVs) of the Higgs fields. The main
ingredient of the model is an approximate global symmetry, which determines
the order of magnitude of Yukawa couplings and scalar VEVs. We assume that
the terms violating this symmetry are suppressed by powers of the small
parameter $\epsilon_{PQ}=O(m_b/m_t)$. The tree-level flavor changing
interactions are small due to a combination of this global symmetry and a
flavor symmetry, but they can be the dominant source of $CP$ violation.
All $CP$-violating effects occur at order $\epsilon_{PQ}^2$ as the result
of exchange of {\it almost}-decoupled extra Higgs bosons and/or through the
usual mechanisms with an {\it almost}-real CKM matrix. On dimensional
grounds, the model gives $\epsilon_K\approx \epsilon_{PQ}^2$ and predicts
for the neutron electric dipole moment (and possibly also for $\epsilon'_K$)
a suppression of order $\epsilon_{PQ}^2$ with respect to the values obtained
in standard and minimal supersymmetric scenarios. The predicted $CP$
asymmetries in $B$ decays are generically too small to be seen in the near
future. The mass of the lightest neutral scalar, the strong $CP$ problem, and
possible contributions to the $Z$ decay into $b$ quarks (the $R_b$ puzzle)
are also briefly addressed in the framework of this model.

\vspace{0.2in}

\end{quotation}
\end{titlepage}

\baselineskip=19pt

\section{Introduction}

Although the standard model is today in impressive agreement with all
particle physics data, its scalar sector has not been proven yet.
A scalar sector defined by elementary fields seems to contradict the
possibility of two very different mass scales (namely, the electroweak
and the grand unification or Planck scales). Supersymmetry (SUSY)
\cite{nill81} would offer an explanation for the stability at
the quantum level of the different scales of the theory, provided that
it is broken only by soft terms below the TeV region. A lot of
attention has been paid to the minimal SUSY extension of the standard
model (MSSM), which presents appealing features such as a consistent
grand unification of the gauge couplings or a candidate for the cold
dark matter of the universe. Experimentally, the MSSM has been so far
{\it flexible} enough to avoid conflict with any measurement, but its
most compelling prediction, the presence of a light neutral Higgs (lighter
than the $Z$ boson at the tree level), is still missing.

The MSSM would also offer distinctive predictions for $CP$ violating
processes. Arbitrary complex phases $\psi$ in soft gaugino masses and
scalar trilinears would give fermion electric dipole moments (EDMs)
well above their present experimental limits. This implies generically
$\psi \leq 10^{-2}$ \cite{deru90}, a somewhat unnaturally small number.
The MSSM would also predict unsupressed flavor changing neutral currents
(FCNCs) unless there is some degree of degeneracy between squark masses
(something which occurs for supergravity Lagrangians with canonical kinetic
terms) and correlation between the Cabibbo-Kobayashi-Maskawa (CKM) matrix
and its equivalent in the squark sector. In the usual SUSY scenario
\cite{bigi90} $CP$ violation in $K$ and $B$ physics depends
essentially on only one phase (the CKM phase $\phi$), whereas
the set of small phases in soft terms (uncorrelated to the family
structure) may have experimental relevance only in fermion EDMs. This
scenario, however, holds only for highly degenerated squark masses.
In general, taking the experimental limit ${ {\Delta m^2_{\tilde{q}} }
\over { m^2_{\tilde{q}} } } \leq {1 \over 30}$ \cite{elli82}
from $K-\bar{K}$ mixing one obtains that acceptable complex
phases in gaugino masses may have an impact on the $K$ system. For
complex gluino masses this was shown in \cite{lang84}, and a model
with small phases has been recently proposed in \cite{babu94a}. It was
also shown \cite{poma93} that due to large top-quark effects acceptable
complex phases in chargino mass terms may also contribute to $CP$
asymetries in the $K$ system.

A different approach to the origin of $CP$ violation which is specially
appealing in SUSY models is the idea of spontaneous $CP$ violation
(SCPV) \cite{tlee73}. All the phases in the Lagrangian (initially
$CP$-conserving) would have their origin in a small number of complex
vacuum expectation values (VEVs) of scalar fields. Moreover, the sizes
of these phases could be correlated by approximate symmetries suppressing
some couplings in the Lagrangian \cite{jliu87}. Unfortunately, in the
minimal Higgs sector of the MSSM there is no room for SCPV
\cite{maek92,poma92}, and {\it hard} $CP$ violation is required.
The possibility of SCPV has been also studied in SUSY extensions with
singlet fields \cite{poma93,babu94b}, where it can be obtained but seems
to require certain ammount of fine tuning.

The SUSY models with more than two Higgs doublets are an obvious
extension of the MSSM \cite{flor83}. They are minimal in the sense
that no new species are introduced, but just repeated. From the model
building point of view there is no compelling reason to disregard
them, and they could appear naturally in models with fermion-Higgs
unification (like $E_6$) \cite{gurs76} or left-right symmetric
scenarios, where two bidoublets are required in order to obtain
realistic fermion masses and mixings. Four Higgs doublet (4HD) models
require an intermediate scale to be consistent with grand unification,
but even this could be more in line with recent data on
$\alpha_s(M_Z)$ than the desert scenario \cite{brah95}. Since more
than one Higgs doublet couples to quarks of a given charge, a possible
concern in this type of models is the presence of FCNC at the tree
level. The experimental limits, however, can be easily avoided just
by invoking the action of an approximate flavor symmetry (see next
section). On the other hand, a nonminimal scalar sector opens the
possibility of SCPV and, in general, widens the parameter space relevant
in low-energy precision measurements (this could be convenient,
for example, if the anomalous value of $R_b$ persists).

In this paper we present a 4HD SUSY model which seems to contain
satisfactory answers to many phenomenological questions. $CP$ violation
appears softly, in complex Higgs masses and VEVs. The main ingredient
of the model is a Peccei-Quinn like approximate symmetry which
determines the order of magnitude of Yukawa couplings and scalar VEVs.
We define this symmetry in such a way that the additional pair of
doublets has small VEVs with order one complex phases and is weakly
coupled to all matter fields. As a consequence, the ratio $m_b/m_t$,
$CP$-violating effects in $K$ physics, and the neutron EDM will
appear suppressed by powers of the small parameter $\epsilon_{PQ}$
that parametrizes the violation of this symmetry. The $CP$ asymmetries
in $B$ decays are predicted to be typically two or three orders of
magnitude smaller than in CKM scenarios (the CKM matrix in the model
is essentially real), a signal that can be used to discriminate this
4HD model with respect to the MSSM or the standard model.

The plan of the paper is as follows. In Section 2 we write the
generic Lagrangian for 4HD SUSY models and review previous
results on spontaneous $CP$ violation. We show that a realistic
scenario for soft $CP$ violation requires complex Higgs masses
in the initial effective model. In Section 3 we define our model and
minimize the Higgs potential. We show that the order of magnitude of
Yukawa couplings, complex scalar VEVs, and the CKM complex
phase are correlated by the approximate global symmetry. In Section 4
we explore the implications of the model on $K$ and $B$ physics as
well as on the neutron EDM. In Section 5 we discuss the resulting
spectrum of scalar fields (in particular, the mass of the lightest
neutral mode) and other possible phenomenological impacts of the 4HD
model. Section 6 is devoted to conclusions. Details about the
minimization of the scalar potential can be found in the Appendix.

\section{ Complex VEVs in four Higgs doublet models }

The most general superpotential with four higgs doublets is given by
\ba
W & = & Q ( {\bf h}_1 H_1 + {\bf h}_3 H_3) D^c +
Q ( {\bf h}_2 H_2 + {\bf h}_4 H_4) U^c +
        L({\bf h}_1^e H_1 + {\bf h}_3^e H_3) E^c \nonumber\\
& + & \mu_{12} H_1 H_2 + \mu_{32} H_3 H_2
+ \mu_{14} H_1 H_4 + \mu_{34} H_3 H_4 ,
\label{eq:superpot}
\ea
where $Q$ stands for quark doublets, $D^c$ for down quark singlets,
$U^c$ for up quark singlets, $L$ for lepton doublets, $E^c$ for
charged lepton singlets, and  ${\bf h}_i$ are the Yukawa matrices
(family indices are omitted). The Higgs doublets $H_1,\;H_3$
and $H_2,\;H_4$ have hypercharges $-1$ and $+1$, respectively.

Including soft SUSY breaking terms the effective potential
for the Higgs fields is
\ba
V & = & m_1^2 H_1^\dagger H_1+ m_2^2 H_2^\dagger H_2
 + m_3^2 H_3^\dagger H_3
+ m_4^2 H_4^\dagger H_4 + \nonumber\\
 & + & (m^2_{12} H_1 H_2 + h.c.) + (m^2_{32} H_3 H_2 + h.c.) + \nonumber\\
 & + & (m^2_{14} H_1 H_4 + h.c.) + (m^2_{34} H_3 H_4 + h.c.) + \nonumber\\
 & + & (m^2_{13} H_1^\dagger H_3 + h.c.)
+ (m^2_{24} H_2^\dagger H_4 + h.c.) + V^{4HD}_D + \Delta V,
\label{eq:pot}
\ea
where $V^{4HD}_D$ contains the D-terms and $\Delta V$ the radiative
corrections. For the neutral components $\phi_i$ of the doublets one has
\ba
V^{4HD}_D & = & \frac {1} {8} (g^2+g'^2)\; [
\phi_1^\dagger\phi_1 +\phi_3^\dagger\phi_3 -\phi_2^\dagger\phi_2 -
\phi_4^\dagger\phi_4 ]^2.
\ea
The radiative contributions $\Delta V$ are generated by SUSY
breaking effects. In our arguments it will suffice to consider
the terms derived from large top (and possibly bottom) quark
Yukawa interactions \cite{elli91}:
\ba
\Delta V & = & \sum\limits_{q} \frac {3} {16 \pi^2} \{
m^4_{\tilde{q}} [ ln({m^2_{\tilde{q}} \over Q^2}) - {3 \over 2} ] -
m^4_q [ ln({m^2_q \over Q^2}) - {3 \over 2} ]  \},
\label{eq:radiative}
\ea
where $q=t,b$,
$m_t^2 = | h_{2t} \phi_2 + h_{4t} \phi_4|^2$,
$m_b^2 = | h_{1b} \phi_1 + h_{3b} \phi_3|^2$,
and $m_{\tilde{q}}^2 = m_{s}^2 + m_q^2$.

Our first comment about the viability of 4HD models should make reference
to the size of FCNCs via Yukawa interactions. If the Yukawa matrices in
(\ref{eq:superpot}) are uncorrelated, there is no reason to expect that
the unitary transformations defining mass eigenstates also diagonalize
(in flavor space) the couplings to the extra Higgs doublets. This would
introduce unsuppressed FCNCs at tree level. The observed pattern of quark
masses and mixings, however, strongly suggests the possibility of an
approximate flavor symmetry as the origin of the hierarchies required in
the Yukawa matrices. In the simplest scenarios \cite{frog79,anta92} the
effect of such a symmetry would be to generate fermion matrices with
off-diagonal elements of order $O(\sqrt{m_im_j}/v)$, where $m_i$ is the
mass of the $i$th quark and $v$ is the weak scale. If the extra Higgs
doublets are a replica (with respect to this flavor symmetry) of the first
doublet, they will introduce Yukawa matrices with the same approximate
structure. In that case the smallness of Yukawa couplings is enough to keep
all FCNC within the experimental limits. In particular, for extra Higgs masses
around 1 TeV the tree-level contributions to $K-\bar{K}$ and $B-\bar{B}$
mixings would be of the same order as the standard contributions
\cite{anta92,hall93}. In our 4HD model we will assume this type of
approximate flavor symmetry at work.

Our main motivation to study 4HD models concerns the origin of
$CP$ violation. In these models explicit $CP$ violation seems even
more inconvenient than in the MSSM, due to new processes mediated
by the Yukawa interactions described above (the approximate flavor
symmetry would not explain, for example, the small value of $\epsilon_K$
\cite{hall93}). The possibility of SCPV in 4HD models has been addressed
in a recent paper \cite{masi95}. There we assume that all the parameters
in the Lagrangian are real and the $CP$-violating phases appear via
VEVs $v_i e^{i\delta_i}$ of the Higgs fields. We showed that at tree level
({\it i.e.}, $\Delta V=0$) the minimum equations for the phases can be
solved in terms of a simple geometrical object but the remaining conditions
are then uncompatible. Namely, after a redefinition of masses and fields
that cancels the terms $m^2_{13}$ and $m^2_{24}$, the four tree-level minimum
conditions for the VEVs $v_i$ with nonzero phases read \cite{masi95}
\ba
v_1 \frac {\partial V} {\partial v_1} & = & v^2_1
     \, [ \, m_1^2
 -{{m^2_{12} m^2_{34} - m^2_{14} m^2_{32}}\over {m^2_{32} m^2_{34}}}
{1\over h({\bf v})}  +  g({\bf v}) \, ] = 0 \nonumber\\
v_2 \frac {\partial V} {\partial v_2} & = &  v^2_2
    \, [ \, m_2^2
    - m^2_{12} m^2_{32} h({\bf v})
  - g({\bf v}) \, ] = 0 \nonumber\\
v_3 \frac {\partial V} {\partial v_3} & = &  v^2_3
    \, [ \, m_3^2
 +{{m^2_{12} m^2_{34} - m^2_{14} m^2_{32}}\over{m^2_{12} m^2_{14}}}
{1 \over h({\bf v})}  + g({\bf v}) \, ] = 0 \nonumber\\
v_4 \frac {\partial V} {\partial v_4} & = &   v^2_4
    \, [ \, m_4^2  + m^2_{14} m^2_{34} h({\bf v})
 - g({\bf v}) \, ] = 0 \, ,
\label{eq:tritvevtwo}
\ea
where $g({\bf v}) =   {1 \over 8} (g^2 + g'^2)
            [ v_1^2 + v_3^2 - v_2^2 - v_4^2 ]$, and
\be
h({\bf v}) = \sqrt{ m^2_{12} m^2_{34} - m^2_{14} m^2_{32}}
\sqrt{ \frac
{ {1 \over {m^2_{32} m^2_{34}}} v^2_1 - {1 \over {m^2_{12} m^2_{14}}} v^2_3 }
{ m^2_{12} m^2_{32} v^2_2 -  m^2_{14} m^2_{34} v^2_4 }  } \, .
\ee
Since the four equations above depend on only two combinations of VEVs,
$g({\bf v})$ and $h({\bf v})$, there will be no solution ({\it i.e.},
phases different from 0 or $\pi$) unless a fine tuned value of the
mass parameters is imposed. Moreover, if this fine tuning were
used it would imply the presence of two massless scalar fields.

The effect of the radiative corrections is twofold: they relax the
ammount of fine tuning required in the equations above, and they
generate masses for the two massless modes. For these two effects to be
sizeable we need (see $\Delta V$ in Eq.~\ref{eq:radiative}) large squark
masses and  at least two large  Yukawa couplings. These could be the two
top quark couplings $h_{2t}$ and $h_{4t}$ or one top ($h_{2t}$) plus one
bottom  ($h_{1b}$) coupling. However, for $m_{s} \leq 5$ TeV and Yukawas
smaller than $\approx 1.2$ (as required to avoid Landau poles before the
Plank scale) we find that the two light scalar fields have masses smaller
than $\approx 30$ GeV. In consequence we conclude that in 4HD models with
all the parameters real the presence of nontrivial complex phases in the
Higgs VEVs implies two scalar fields apparently too light. (A more detailed
examination of the parameter space might show, however, that this possibility
is not entirely excluded by current limits on the masses of the scalar
fields.) The situation here is then similar to the MSSM (where the allowed
mass of the light scalar field is already excluded \cite{poma92}) or the
singlet model (which relay on radiative effects to give mass to a mode with
negative tree-level mass \cite{babu94b}).

In 4HD scenarios there is, however, still another possibility which seems
consistent with the idea of SCPV. It requires that the four Higgs mass
parameters $\mu$ in the superpotential (or, equivalently, the six parameters
$m^2_{ij}$ in Eq.~\ref{eq:pot}) are allowed to be complex. This could be
justified since the Higgses are the only superfields which are not protected
of mass contributions by the gauge symmetry. They could acquire their masses
in an intermediate scale, via complex VEVs of singlet fields with no sizeable
effect on the rest of the low-energy effective Lagrangian. We will not assume
complex phases on all soft SUSY-breaking terms, since in principle these
singlets do not couple to gauginos or squarks (we will neglect the possibility
of further phases or new contributions to SUSY-breaking parameters due to the
presence of nonsinglet heavy fields \cite{dann85}). The hypothesis of complex
Higgs masses, consistent with a {\it soft} origin of $CP$ violation, is not
possible in the MSSM or the singlet model, since there (unlike here) all
the Higgs masses can be made real by field redefinitions.

In the next sections we study the implications of a 4HD model where all the
parameters in the initial Lagrangian are real except for the Higgs mass
parameters.

\section{ Definition of the model }

The approximate flavor symmetry described in the previous section
suppresses all FCNC amplitudes to acceptable limits. However,
multi-Higgs models face a potential problem also with $CP$ violation:
if the Yukawa couplings are complex with phases of order one,
$CP$-violating signals in $K$ physics would be too large. In particular,
$\epsilon_K$ would be typically two or three orders of magnitude larger
than observed \cite{hall93}. Thus it seems that a general model of many
Higgs doublets requires another ingredient in addition to the flavor
symmetry. Its effect should be either a suppression of the Yukawa
couplings of the new doublets, or to make the complex phases small.
The first approach is typical in models with natural flavor conservation
(NFC) \cite{glas77}, whereas a natural suppression of the phases has been
obtained in the superweak model with SCPV proposed in \cite{jliu87}.
In our scenario these two effects will be achieved by the action of a
global symmetry.

We will assume that the effective Lagrangian of the model obeys an
approximate Peccei-Quinn like symmetry with the following assignment
of charges \cite{rasi95}:
\be
Q(H_3) =  +1\;\;\;Q(H_4)  =  -1\;\;\;Q(D^c)  = +1\;.
\ee
All other superfields have zero charge\footnote{ There should also be
charge assignments in the lepton sector, {\it e.g.} $Q(E^c) = +1$.
We will comment on this possibility when discussing the electron EDM.}.
The symmetry is approximate in the sense that couplings of operators
violating the symmetry are suppressed by powers of a small parameter
$\epsilon_{PQ}$. In this section we discuss the impact of this global
symmetry first on the Higgs scalar part of the potential and then on the
Yukawa sector.

The assignment of charges tells us that in the scalar potential $m_{14}^2$,
$m_{32}^2$, $m_{13}^2$, $m_{24}^4$ are suppressed
by $\epsilon_{PQ}$. $m_{12}^2$ and $m_{34}^2$ remain unsuppressed (of the
order of the SUSY-breaking scale $m_s\leq 1$ TeV). For easy reading we will
write the suppression factors explicitly; for example $m^2_{32}$ becomes
$\epsilon_{PQ} m^2_{32}$, where $m^2_{32}=O(m^2_s)$. The tree-level scalar
potential (we neglect radiative corrections in the following)
involving only neutral Higgs fields is then given by
\ba
V & = &
\left (
\begin{array}{cc} \phi_1^\dagger  &  \phi_3^\dagger   \\
\end{array} \right ) \left (
\begin{array}{cc}
m_1^2 &  \epsilon_{PQ} m_{13}^2 \\
\epsilon_{PQ} m_{13}^{2*} &  m_{3}^2 \\
\end{array} \right ) \left (
\begin{array}{c} \phi_1 \\  \phi_3 \\
\end{array} \right ) \nonumber\\
 & + & \left (
\begin{array}{cc} \phi_2^\dagger  &  \phi_4^\dagger   \\
\end{array} \right ) \left (
\begin{array}{cc}
m_2^2 &  \epsilon_{PQ} m_{24}^2 \\
\epsilon_{PQ} m_{24}^{2*}  &  m_{4}^2 \\
\end{array} \right ) \left (
\begin{array}{c} \phi_2 \\  \phi_4 \\
\end{array} \right ) \nonumber\\
 & + & [ \left (
\begin{array}{cc} \phi_1 &  \phi_3 \\
\end{array} \right ) \left (
\begin{array}{cc}
m_{12}^2 &  \epsilon_{PQ} m_{14}^2 \\ \epsilon_{PQ} m_{32}^2 &  m_{34}^2 \\
\end{array} \right ) \left (
\begin{array}{c} \phi_2 \\  \phi_4 \\
\end{array} \right ) + h.c. ]\nonumber\\
 & + &  {1 \over 8} (g^2+g'^2) [
\left (
\begin{array}{cc} \phi_1^\dagger  &  \phi_3^\dagger   \\
\end{array}
\right ) \left (
\begin{array}{c} \phi_1 \\  \phi_3 \\
\end{array}
\right ) - \left (
\begin{array}{cc} \phi_2^\dagger  &  \phi_4^\dagger   \\
\end{array}
\right ) \left (
\begin{array}{c} \phi_2 \\  \phi_4 \\
\end{array} \right ) ]^2 \, .
\label{eq:potapprox}
\ea

As explained in the previous section, we assume that the mass parameters
$m^2_{ij}$ are complex. The first two mass matrices above can be
diagonalized through two unitary transformations of order $\epsilon_{PQ}$
of the scalar fields:
\be
\left (
\begin{array}{c} \phi'_1 \\  \phi'_3 \\
\end{array} \right ) = {\bf U}_1
\left (
\begin{array}{c} \phi_1 \\  \phi_3 \\
\end{array} \right ); \;\;
\left ( \begin{array}{c}
\phi'_2 \\  \phi'_4 \\
\end{array}
\right ) = {\bf U}_2 \left (
\begin{array}{c} \phi_2 \\  \phi_4 \\
\end{array} \right ).
\label{eq:unit}
\ee
The quartic term in the potential will not change its form and
can be obtained just by replacing unprimed by primed fields. The
relative size of the four complex masses in the third mass matrix
above will stay the same ({\it i.e.}, the off-diagonal elements are
still suppressed by $\epsilon_{PQ}$). A phase transformation of the
fields $\phi_i$ can be used to remove three of the four phases,
leaving only one phase $\alpha$ in the scalar potential. Dropping
the prime to specify transformed quantities, the mass terms read
\ba
V_m & = &
\left (
\begin{array}{cc} \phi_1^\dagger  &  \phi_3^\dagger   \\
\end{array} \right ) \left (
\begin{array}{cc}
m_1^2 &  0 \\ 0 &  m_{3}^2 \\
\end{array} \right ) \left (
\begin{array}{c} \phi_1 \\  \phi_3 \\
\end{array} \right ) + \left (
\begin{array}{cc} \phi_2^\dagger  &  \phi_4^\dagger   \\
\end{array} \right ) \left (
\begin{array}{cc}
m_2^2 &  0 \\ 0 &  m_{4}^2 \\
\end{array} \right ) \left (
\begin{array}{c} \phi_2 \\  \phi_4 \\
\end{array} \right ) \nonumber\\
 & + & [ \left (
\begin{array}{cc} \phi_1 &  \phi_3 \\
\end{array} \right ) \left (
\begin{array}{cc}
m_{12}^2 &  \epsilon_{PQ} e^{i\alpha} m_{14}^2 \\
\epsilon_{PQ} m_{32}^2 &  m_{34}^2 \\
\end{array} \right ) \left (
\begin{array}{c} \phi_2 \\  \phi_4 \\
\end{array} \right ) + h.c. ]\; .
\ea
where the masses are now real and the phase $\alpha$ of $m^2_{14}$
has been written explicitly. We assume $\alpha$ to be of order
one, since there is no symmetry reason for it to be suppressed.
It is easy to see how the Higgs field redefinitions above change
the mass parameters $\mu_{ij}$ in the superpotential $W$ (these
parameters are relevant since they will appear in scalar trilinears).
We obtain $\mu_{12}$ and $\mu_{34}$ real up to order $\epsilon_{PQ}^2$
whereas $\mu_{14}$ and $\mu_{23}$ will be mass coefficients
with arbitrary complex phases but suppreessed by a power of
$\epsilon_{PQ}$.

We now go to the minimization of the Higgs potential. In particular,
we want to find what are the relative size and the phases of the
scalar VEVs suggested by the approximate symmetry. We write
\be
<\phi_1> = {1 \over \sqrt{2} } v_1  \; ; \;\;
<\phi_3> = {1 \over \sqrt{2} } v_3 e^{i\delta_3} \, ,
\ee
and
\be
<\phi_2> = {1 \over \sqrt{2} } v_2 e^{i\delta_2}  \; ; \;\;
<\phi_4> = {1 \over \sqrt{2} } v_4 e^{i\delta_4} \, ,
\ee
where a global hypercharge transformation has been used to rotate away the
phase of $<\phi_1>$. A detailed discussion of the minimum equations can be
found in the Appendix. The results are the following. For nonzero values of
the phase $\alpha$ the minimum is allways complex. The suppression in terms
of $\epsilon_{PQ}$ of the mass parameters determines the order of magnitude
of the VEVs and phases:
\ba
v_1 \, ,\; v_2 & = & O(v) \nonumber \\
v_3 \, ,\; v_4 & = & O(\epsilon_{PQ} v) \nonumber \\
\delta_2 & = & O(\epsilon^2_{PQ}) \nonumber \\
\delta_3 \, ,\; \delta_4 & = & O(1)
\label{eq:structure}
\ea
where $v$ denotes the weak scale. A remarkable feature of the model is that
one can understand its structure in terms of an expansion in $\epsilon_{PQ}$
from the model with just two doublets. In the limit $\epsilon_{PQ}=0$
the sectors $(H_1,H_2)$ and $(H_3,H_4)$ decouple; the minimum gives
then equations for $v_1$ and $v_2$ identical to the VEVs in the MSSM,
whereas $v_3=v_4=0$. The phase $\delta_2$ is then zero, while the
phases $\delta_3$ and $\delta_4$ are irrelevant. Turning on a small
value of $\epsilon_{PQ}$ gives (proportionally) VEVs to the extra pair
of scalars. Simultaneously it allows a nonzero value of $\alpha$,
which translates into unsuppressed complex phases in the $(H_3,H_4)$ sector
and a phase of order $\epsilon^2_{PQ}$ in the $(H_1,H_2)$ {\it standard}
sector. The mixings between the two sectors are small: either in a basis of
scalar mass eigenstates or in a basis where $v'_3=v'_4=0$ (useful when
discussing FCNCs via Yukawas), both are obtained from the original basis
just by unitary transformations of order $\epsilon_{PQ}$.

We now turn to the Yukawa sector of the theory. The charge assignments
dictates that the matrix ${\bf h}_2$ is unsuppressed, ${\bf h}_1$ and
${\bf h}_4$ are suppressed by a factor of $\epsilon_{PQ}$, while
${\bf h}_3$ is suppressed by $\epsilon_{PQ}^2$. Making this
suppression explicit the Yukawa sector for the quark fields reads
\ba
{\cal L}_Y & = & Q ( \epsilon_{PQ} {\bf h}_1 H_1 +
            \epsilon_{PQ}^2 {\bf h}_3 H_3) D^c +
           Q ( {\bf h}_2 H_2
            + \epsilon_{PQ} {\bf h}_4 H_4) U^c \;,
\label{eq:superpotsup}
\ea
where now ${\bf h}_i$ ($i=1,...,4$) just carry the suppression from
the flavor symmetry. In our model the Yukawa couplings of the initial
Lagrangian are real. However, the field redefinitions performed to leave
only one phase $\alpha$ in the Higgs potential will also redefine the
Yukawa couplings and introduce complex phases. As we show, these phases
translate into a CKM phase and complex FCNC couplings which are
naturally suppressed by the approximate global symmetry.

We first performed the unitary transformations in Eq.~(\ref{eq:unit}),
which redefine the fields $H_i$ by (complex) factors of order
$\epsilon_{PQ}$. They imply a redefinition of the Yukawa matrices which
introduces phases of order $\epsilon_{PQ}^2$ in ${\bf h}_1$ and ${\bf h}_2$
and of order one in ${\bf h}_3$ and ${\bf h}_4$. Then we performed the
(order one) phase redefinitions of the Higgs doublets that make all mass
parameters real except for $m^2_{14}$. This translates into overall phases
of order one multiplying the Yukawa matrices ${\bf h}_i$. However, we can
still redefine the quark fields and absorb the phases which multiply
${\bf h}_1$ and ${\bf h}_2$ (the leading Yukawa couplings). The net result
is that the Lagrangian in (\ref{eq:superpotsup}) expressed in terms of the
Higgs fields used to minimize the potential has real (up to order
$\epsilon_{PQ}^2$) couplings in ${\bf h}_1$ and ${\bf h}_2$ and arbitrary
complex phases in ${\bf h}_3$ and ${\bf h}_4$.


After spontaneous symmetry breaking, the structure of VEVs in
(\ref{eq:structure}) suggests (note that $v_1$ amd $v_2$ are not suppressed
by powers of $\epsilon_{PQ}$) $\tan \beta\equiv \sqrt{v_2^2+v_4^2\over
v_1^2+v_3^2}=O(1)$ and $\epsilon_{PQ} = O (m_b / m_t)$. Thus, the
approximate global symmetry is used to accommodate the small ratio
$m_b/m_t$, while the hierarchy between generations of the same charge is
left to the flavor symmetry (the flavor symmetry would be exact in the
limit with only the third generation being massive). The complex Yukawas
${\bf h}_3$ and ${\bf h}_4$ and the corresponding VEVs ($v_3$ and $v_4$)
are both suppressed by a power of $\epsilon_{PQ}$, whereas the leading
Yukawas and VEVs are real up to order $\epsilon_{PQ}^2$. We then obtain
quark mass matrices where all the entries have complex phases of order
$\epsilon_{PQ}^2$. In consequence, the complex phase in the CKM matrix is
also of order $\epsilon_{PQ}^2$.

It is also easy to see what is the pattern of FCNC and $CP$ violation via
scalar exchange predicted by the model. It will be convenient to define
a basis where only two of the four Higgs fields develop VEV and then only
the second pair of scalars mediates FCNC processes. Again, this involves a
unitary transformation of order $\epsilon_{PQ}$, which leaves almost
decoupled the extra pair of Higgses (essentially ($H_3,H_4$)). In addition,
the mixings in the scalar mass matrices between the two sectors are also
suppressed. The overall suppression by a power of $\epsilon_{PQ}$,
when added to the one with origin in the flavor symmetry, renders these
tree-level FCNCs smaller than CKM (box) diagrams typically by a factor
of $\epsilon^2_{PQ}$. In particular, the $K-\bar{K}$ and $B-\bar{B}$ mixings
are dominated here by the standard contributions, like the $W$-exchange box
diagram in Fig. \ref{fig:wexchange}. This fact will distinguish our scenario
from typical multi-Higgs models with soft $CP$ violation where the tree-level
superweak interactions are the main source of flavor changing processes
\cite{jliu87}. On the other hand, the sector ($H_3,H_4$) involves arbitrary
phases in Yukawa couplings and scalar VEVs. Although suppressed by a factor
of $\epsilon^2_{PQ}$, these couplings can be the dominant source of $CP$
violation through diagrames like the one shown in Fig.~\ref{fig:treelevelfc}.
In particular, as shown in the next section, they are the main source of
complex phases in $K$ physics and compete with box contributions in $B$
physics.

We need as well that FCNC contributions via SUSY particles (wino and
gluino box diagrams) are within the experimental limits, which in general
requires certain degree of squark-quark alignment and squark degeneracy.
In fact, the squark-quark alignment could appear here as a natural
consequence of the flavor symmetry\cite{ynir93}. Since we assumed no complex
phases in soft-SUSY parameters others than Higgs masses, their $CP$ violating
effects will follow the same pattern described above. We explore these and
other phenomenological implications of the model in the next sections.

\section{ $CP$ violation in $K$ and $B$ physics and the neutron EDM }

{\it $K$ physics.}
As our first example we look at the $K$ system. In this scenario the
FCNC processes via Yukawa interactions (see previous section)
are highly suppressed, and the dominant contribution to Re $\Delta M_{12}$
comes from the box diagram in Fig. \ref{fig:wexchange}. The leading imaginary
contribution to $\Delta M_{12}$, however, will come from the neutral Higgs
exchange in Fig. \ref{fig:treelevelfc}. The flavor-changing Yukawa couplings
are complex (with phases of order 1) and generically suppressed by the
Peccei-Quinn and the flavor symmetries (for example, in
Fig. \ref{fig:treelevelfc} the couplings are of order $\epsilon_{PQ}
\sqrt{m_d m_s} / v$). When the mass of the exchanged scalar is around 1 TeV,
this (complex) diagram is roughly suppressed by $\epsilon^2_{PQ}$ with
respect to the box diagram in Fig.~\ref{fig:wexchange}. Since the CKM matrix
and then the box contributions are approximately real, the leading
contribution to Im $(\Delta M_{12})$ comes from the Higgs exchange in
Fig. \ref{fig:treelevelfc}. On dimensional grounds, the $CP$ violating
parameter $\epsilon_K$ (see \cite{chau83} for definitions and notation)
in the $K$ system and $\epsilon_{PQ}$ are related:
\be
|\epsilon_{K}| \approx {1\over 2\sqrt{2}}
\frac { {\rm Im}\; \Delta M_{12}} { {\rm Re}\; \Delta M_{12} }
\approx \epsilon_{PQ}^2.
\ee
The parameter $\epsilon_{PQ}$ sets the overall strength of Yukawa couplings
of the Higgs doublets and suggests the order of magnitude of all the scalar
VEVs. In particular (see section 3), one expects $\tan\beta= O(1)$ and
$\epsilon_{PQ}=O(m_b/m_t)$. Then the relation above establishes
$\epsilon_K\approx 10^{-3}$, as experimentally required.

Other sizeable contributions to $\epsilon_K$ may come from
SUSY box diagrams with chargino or gluino exchange. Both of them
require large SUSY contributions to $\Delta M_{12}$ (of the same
order as the standard box diagram). Chargino box diagrams
would then give \cite{poma93} contributions of order $\epsilon_K\approx
10^{-1}\epsilon_{PQ}^2$, whereas gluino boxes \cite{babu94a}
could be as large as $\epsilon_K\approx \epsilon_{PQ}^2$
({\it i.e.}, of the same order as the dominant tree-level scalar
exchange). The factors $\epsilon_{PQ}^2$ above derive from the
suppresion in Yukawa couplings or extra scalar VEVs. Large gluino
box contributions, however, also require large left-right
squark mixing $\delta_{LR}\equiv m^2_{LR}/m^2_s\approx 10^{-3}$
(a naive estimate would give $\delta_{LR}\approx
{A\sqrt{m_dm_s}\over m_s^2}\approx 10^{-4}$). In addition, the
tree-level contributions to $\epsilon_K$ can be easily enhanced
\cite{rasi95} assuming Higgs masses ligheter than 1 TeV, so
the clear tendency in our model is that this type of nonstandard
Higgs exchange provides the dominant contribution to $\epsilon_K$.

In contrast, the expected value for $\epsilon'_K$
differs in principle from the standard
model prediction. An estimate of $\epsilon'_K$ can be obtained
from the phase $t_0\equiv$ Im$A_0$/Re$A_0$, where $A_i$ is
the decay amplitude of a $K^0$ into two pions of isospin $i$
(see \cite{chau83} for notation). In particular
one has the experimental constraint
\be
t_0 \approx \sqrt{2} |{A_0\over A_2}| |\epsilon'_K|
\leq 10^{-4}\;.
\ee
The dominant contribution to
Re$A_0$ arises from the standard penguin diagram:
\be
{\cal L}_p\approx {\alpha_s\alpha_W\over 3m_W^2}
\sin \theta_c \ln {m_c^2\over m_K^2}{\cal O_{LR}}+H.c.\;,
\ee
where
\be
{\cal O_{LR}} =
(\bar{s}_L\gamma_\mu T^a d_L)
(\bar{q}_R\gamma_\mu T^a q_R)
\ee
and $T^a$ are the 3-dimensional generators of $SU(3)$. In our scenario,
however, the imaginary part of this penguin diagram is suppressed by the
smallness of the phase (of order $\epsilon_{PQ}^2\approx 10^{-3}$) in
the CKM matrix. Since the standard model prediction for $t_0$ is of order
$s_{13}s_{23}/s_{12}\approx 10^{-3}$, we obtain a first contribution of
order $10^{-6}$. Other contributions to $t_0$ may come from
penguin diagrams with chargino and stop (Figure \ref{fig:charginoep})
and tree-level diagrams with charged scalars.
The first contributions have been studied in \cite{poma93} in the
context of SUSY models with SCPV. It is found that they are typically
of order $t_0\approx 10^{-3}\sin\delta$, being
$\delta$ the complex phase in the VEV of $H_2$ (the scalar field
giving mass to the top quark). In our model the Higgs fields
$H$ and $\tilde{H}$ in Fig.~\ref{fig:charginoep} can correspond to $H_2$ or
$H_4$. In the first case the phase $\delta$ is of order
$\epsilon_{PQ}^2$, and in the second case the same degree of
suppression comes from the small VEV and the small Yukawa
coupling of $H_4$. The contributions via exchange of charged
Higgses have been analyzed in \cite{jliu87} in the context of
two-Higgs doublets models. In our model either the Yukawas
are {\it almost} real (for the doublets with dominant couplings, as
it happens in \cite{jliu87}), or the Yukawas themselves are suppressed.
Hence, from the three types of diagrams we obtain contributions
(taking $\epsilon^2_{PQ}\approx 10^{-3}$) of order
$t_0\approx 10^{-6}$ or ${\epsilon'_K\over \epsilon_K}\approx 10^{-5} \;$
(for $|A_2/A_0|\approx 1/22$).

Potentially larger contributions to $\epsilon'_K$ are expected
from gluino-mediated penguin diagrams (Figure \ref{fig:gluinoep}).
Although gluino masses in our model are real, there will be complex
phases of order $\epsilon_{PQ}^2$ (see discussion of chargino penguin
above) in left-right squark mixings.
This type of contributions to $\epsilon'_K$ have been
studied in detail in \cite{babu94a}. There it is found that
for complex phases in gluino masses of order $10^{-3}$
they could result in values as large as ${\epsilon'_K \over \epsilon_K}
\approx 10^{-3}$. An analogous result has been recently obtained
in \cite{poma95}, with contributions from squark mixings of CKM type
and small $CP$-violating phases (which are natural in our scenario).
These values are only obtained, however, when gluino box diagrams
saturate the value of $\epsilon_K$. Since we have assumed that the FCNCs
are here dominated by standard box diagrams, we expect a value for
$\epsilon'_K$ typically smaller. Modulo hadronic matrix element
uncertainties, we estimate
\be
{\epsilon'_K \over \epsilon_K}
\approx 10^{-4} - 10^{-5}\;,
\ee
with the possibility to consistently increase this value via
gluino penguin contributions.

\vskip 0.5truecm

{\it $B$ physics.}
We consider now $CP$ violation in $B$ physics \cite{ynir92}. Although
today the only observed $CP$ violation is in the $K$ system, the
standard model predicts clear signals in $B$ decays that should be observed
in the near future. These $CP$ asymmetries are generally parametrized in
terms of the complex phases $\lambda_{iq}$, which in turn depend
on the product of phases in three amplitudes: the direct $b$ decay, the
$B-\bar{B}$ mixing, and (possibly) the $K-\bar{K}$ mixing. In CKM scenarios
the phases $\lambda_{iq}$ are constrained by unitarity and have a simple
geometrical interpretation (these predictions are not expected to change
much in minimal SUSY models). However, in our 4HD scenario the three
amplitudes above have complex phases of order $\epsilon^2_{PQ}$ (see
below), and the predictions change to the extent that no $CP$ asymmetries
will be observed at the projected $B$ factory at SLAC.

To see why this is so, we will first consider the decay amplitude of a
$b$ quark into lighter flavors. The main contribution corresponds to a
tree-level diagram with $W$ exchange. Since it will be proportional to
elements of the CKM matrix, its imaginary component will be suppressed
by a factor of $\epsilon_{PQ}^2$. The decay via charged Higgs are suppressed
by the same factor due to the relative smallness of their Yukawas and the
smallness of the mixing (in the scalar mass matrix) between the standard and
the extra Higgs sectors. In non-SUSY models with NFC the second effect can
give significant contributions (proportional to $m_t$) in $B$ decays and
in flavor-changing processes \cite{jliu87}.

The main contribution in this model to $B-\bar{B}$ mixing
$\Delta M_{B\overline B}$  comes from the standard box
diagrams, and is proportional to CKM elements. The tree-level
diagrams with exchange of neutral scalar give contributions which are
suppressed by the flavor and the Peccei-Quinn symmetries, with a
relative factor of $\epsilon_{PQ}^2$ with respect to the box diagrams. Both
types of diagrams introduce imaginary components of order $\epsilon_{PQ}^2$
with respect to the main real component, and the contribution to the complex
phase $\lambda_{iq}$ from $B-\bar{B}$ mixing is negligible (of that order).
As discussed above, the same conclusion applies to the contribution
from $K-\bar{K}$ mixing.

In consequence, in this scenario one expects that all $CP$
asymmetries in $B$ decays negligibly small (of order
$\epsilon^2_{PQ} \approx 10^{-3}$). This type of prediction is shared,
for example, by non-SUSY multi-Higgs doublet models \cite{jliu87,hall93} or
SUSY models with real Yukawa matrices \cite{poma93,babu94a}.
In CKM scenarios the situation is essentially different. There
the smallness of $CP$ violation in the $K$ system is atributed to the
smallness of the CKM elements involving the third family of quarks,
whereas $CP$-violating asymmetries in the $B$ system are large: the
$B-\bar{B}$ mixing and the $b$ decays are proportional to elements of
the CKM matrix with arbitrary complex phases. The absence of $CP$
asymmetries at the SLAC $B$ factory would point to a non-CKM origin of
$CP$ violation, and many-Higgs doublet model (SUSY or non-SUSY) would appear
as a natural candidate.

\vskip 0.5truecm

{\it Neutron EDM:}
As in usual SUSY scenarios, the prediction of our model for
the neutron EDM $d_n$ is much larger than in the non-SUSY standard
model. In the MSSM the explicit phases $\psi$ in SUSY-breaking
gaugino mases and scalar trilinears introduce contributions
which roughly require a suppression of 2 or 3 orders of magnitude:
$d_n \approx 10^{-25} ({ \psi \over {7\times10^{-3}}})\;e$ cm
\cite{deru90}. These diagrams are also present in our scenario, but their
contribution has a natural suppression of order $\epsilon^2_{PQ}$
respect to the MSSM. The origin of this factor is (again!)
either the smallnes of the complex phase $\delta_2$ (of order
$\epsilon_{PQ}^2$) in the two standard Higgs doublets, or the combined
relative smallness of VEVs and Yukawa couplings (both suppressed by
a factor of $\epsilon_{PQ}$) of the two extra doublets.

To illustrate this fact, let us consider the contribution from the
the one-loop chargino-squark diagram (Figure \ref{fig:charginonedm}).
When $H$ corresponds to $H_1$, then the complex phase in the
Yukawa coupling is suppressed. When $H$ is $H_3$, then the VEV and the Yukawa
coupling are of order $\epsilon_{PQ}$. In consequence, in this 4HD
one expects
\be
d_n \approx 10^{-23} \epsilon_{PQ}^2\;e\;{\rm cm}
\approx 10^{-26} \;e\;{\rm cm}\;,
\ee
a value which is close to the present experimental limit
$|d_n|<1.2\times10^{-25}e$ cm \cite{smit90}.

Here we also comment on the lepton sector. We still have the freedom
to assign a global symmetry charge to $E^c$ (or even $L$). For simplicity
let us consider $E^c = +1$, which would be consistent with
$m_\tau=O(m_b)$. In this sector all FCNC processes via
nonstandard scalars will be completely negligible (the size of
Yukawas suggested by the flavor symmetry would be enough to control all
these processes). The pattern of $CP$ violation will be analogous
to the one discussed in the quark sector, with the relevant complex
phases suppressed by a factor of $\epsilon^2_{PQ}$. The leading
contribution to the electron EDM comes from a diagram similar
to the one shown in Fig.~\ref{fig:charginonedm}. If all the
superparticles have comparable masses it is expected
that $d_e \approx 10^{-2} d_n$ \cite{bern91}, so that in our model
the electron EDM is not far from the present experimental limits.

\section{ Other phenomenological implications }

As shown by Flores and Sher in \cite{flor83}, the presence of a light
neutral scalar field (with a tree-level mass smaller than $M_Z$) is a
prediction shared by all SUSY models with Higgs doublets only, regardless
of the number of doublets. Since in the limit $\epsilon_{PQ}\rightarrow 0$
the scalar sector of our 4HD model essentially coincides with the MSSM,
we expect small corrections to the standard predictions.

To see how these corrections arise we will first consider the model with
$\alpha=0$ and, consequently, with all the VEVs real. The approximate
symmetry dictates that $v\approx v_1\approx v_2$ and
$\epsilon_{PQ} v\approx v_3\approx v_4$. We can perform two rotations of
order $\epsilon_{PQ}$ of the Higgs fields (one in the space
$\phi_1-\phi_3$ and another in $\phi_2-\phi_4$) in such a way that
$v_3=v_4=0$. It is then straightforward to find the mass $4\times 4$
matrix $M^2_h$
\be
M^2_h=\left ( \begin{array}{cc}
M^2_0 & M^2_1 \\
M^{2\;T}_1 & M^2_2 \\
\end{array} \right )
\label{eq:mass}
\ee
for the $CP$-even scalar fields $h_i$.
The $2\times 2$ matrix $M^2_0$ corresponding to $h_1-h_2$ is
identical to the one obtained in the MSSM, with an eigenvalue
 $m^2_h$ smaller than $M^2_Z$ and another $m^2_H\approx m^2_{s}$.
The submatrix $M^2_2$ in the $h_3-h_4$ sector has two eigenvalues of order
$m^2_s$. The Peccei-Quinn symmetry forces all the elements in
$M^2_1=O(\epsilon_{PQ} m^2_s)$ and, through mixing, tends to lower the
lightest eigenvalue in $M^2_h$ by terms of order $\epsilon^2_{PQ} m^2_s$.
For nonzero values of $\alpha$ the scalar VEVs will be allways
complex (see Section 3), introducing mixing betweeen CP-odd and
$CP$-even states. Due to the approximate symmetry, however, the mixings
of the lightest scalar field with $CP$-odd states are small and
introduce corrections of the same order. In consequence, we conclude that
these corrections do not change significantly the tree-level bound
$m_h<M_Z$ (for $\epsilon^2_{PQ}=10^{-3}$ and $m_s=500$ GeV this bound
is lowered by less than 2 GeV). However, we expect radiative top quark
effects to be much more important.

We note that the spontaneous breaking of the (approximate) global
symmetry do not introduce light fields. The reason for this is
that in the limit of exact symmetry ($\epsilon_{PQ}=0$) the only VEVs
breaking the symmetry ($v_3$ and $v_4$) go to zero too: there are no
light pseudo-goldstone states because the size of the spontaneous and the
explicit symmetry breaking terms is of the same order.

It is also easy to see that this model accommodates the small
ratio $m_b/m_t$ without need of fine tuning to avoid too light
charginos \cite{nels93} (in the MSSM, a small mass ratio $m_b/m_t$
based on a large value of $\tan\beta$ implies such fine-tuning
problem). The chargino mass matrix is here
\ba
\left (
\begin{array}{ccc}
\mu_{12} &  \epsilon_{PQ} \mu_{14} & {{g v} \over {\sqrt{2}}} \\
\epsilon_{PQ}\mu_{32} &\mu_{34}&{{g \epsilon_{PQ}v}\over{\sqrt{2}}} \\
{{g v} \over {\sqrt{2}}}& {{g \epsilon_{PQ} v}\over{\sqrt{2}}} & M \\
\end{array} \right )\;,
\ea
where we used the VEVs in (\ref{eq:structure}) and denoted the
gaugino mass by $M$. This structure has no light eigenvalues.

Another possible implication of 4HD models concerns the
value of $R_b$. Within the standard model, the partial
width of the $Z$ boson to $b\overline b$ seems to be very
sensitive to top-quark radiative correction. For the top observed
in CDF the predicted value is well below (a three-$\sigma$
deviation) the present experimental limits \cite{pdg94}.
In minimal SUSY scenarios the main correction results from the
balance between $Z$ vertices with Higgs-top and their SUSY partners,
and the anomalous value of $R_b$ can be alleviated for
light charginos and/or light stop scalars \cite{sola95}.
In 4HD SUSY models the situation is similar (especially in our scenario,
due to the global approximate symmetry assumed in the Yukawa sector), with
more freedom than in the MSSM to adjust the corrections. Note, for example,
that large bottom Yukawa couplings do not imply necessarily a large value of
$\tan\beta$ $(\equiv \sqrt{v_2^2+v_4^2\over v_1^2+v_3^2})$.

Our last comment concerns the strong $CP$ problem. In the model under
consideration there are (tree-level) contributions to $\theta$ of order
$\epsilon_{PQ}^2\approx 10^{-3}$, a value much bigger than the present
experimental limit ($\theta<10^{-9}$). It seems possible, however, that
the intermediate scale used to break $CP$ would also define a realistic
axion scenario. Some of the ingredients (a Peccei-Quinn symmetry or
singlet VEVs breaking the global symmetries) are already present in the
model. Of course, for this scenario to work other requirements (on the
dimension of the operators breaking the anomalous Peccei-Quinn symmetry,
on the ratio of the scales involved,...) are also needed.

\section{Conclusions}

The origin of $CP$ nonconservation in SUSY models provides a good reason to
explore nonminimal extensions. In the usual MSSM scenario $CP$-violating
phases occur in two different sectors: in Yukawa couplings, where they would
be responsible for $CP$ violation in $K$ and $B$ physics, and in
SUSY-breaking terms (gaugino masses and scalar trilinears), where they would
induce too large fermion EDMs unless suppressed by two or three orders of
magnitude.

We have presented here an extension of the MSSM with four Higgs doublets
where the complex phases appear spontaneously, induced by explicit phases
in Higgs masses. An approximate Peccei-Quinn symmetry {\it almost} decouples
the pair of extra Higgs fields, but their small couplings (also suppressed
by the flavor symmetry) turn out to be responsible for all $CP$-violating
phenomena. In particular, tree-level FCNC diagrams are irrelevant in
Re$\Delta M_{12}$ but responsible for $\epsilon_K$. The resulting
CKM matrix of the model has a negligible complex phase of order
$\epsilon_{PQ}^2\approx 10^{-3}$. This suppression appears in all $CP$
signals either from small phases in the dominant scalar sector or from
small ratios of VEVs and Yukawa couplings in the extra sector.

On dimensional grounds, the parameter $\epsilon_{PQ}$ specifying the
violation of the Peccei-Quinn symmetry sets:

$\bullet$ the ratio $m_b/m_t \approx \epsilon_{PQ}$, and the relative
suppression of the Yukawa couplings of the extra Higgses
(${\bf h}_3/{\bf h}_1\approx{\bf h}_4/{\bf h}_2\approx \epsilon_{PQ}$);

$\bullet$ the parameter $\epsilon_K \approx \epsilon_{PQ}^2$ and the ratio
$\epsilon'_K/\epsilon_K \le \epsilon_{PQ}^2$ (with a preferred
value ($10^{-1}-10^{-2})\epsilon_{PQ}^2$);

$\bullet$ the neutron EDM $\approx 10^{-23}\epsilon_{PQ}^2 e$ cm,
being $10^{-23}$ an estimate for typical SUSY-breaking parameters;

$\bullet$ and the $CP$ violating asymmetries $\lambda_{iq}
\approx \epsilon_{PQ}^2$ involved in $B$ physics.

In consequence, a neutron EDM close to its present experimental
limit, negligible $CP$-violating effects on $B$ physics,
and a small value of the $\epsilon'_K$ parameter could be regarded as
typical predictions of the model. In addition, we have estimated the
effects of the extra sector on the mass of the lightest neutral scalar
and commented on other aspects of the model ($R_b$ and the strong
$\theta$ parameter).

We think that 4HD models constitute an interesting possibility in
SUSY extensions which, however, seems almost absent in the
literature. We have defined a scenario where $CP$ violation
is brought under control in a consistent way (due to the action of
an approximate symmetry), in contrast to SUSY models where the complex
phases are assumed small without explanation. Although we have analyzed
a particular model, we think that it contains essential ingredients which
may be shared by any satisfactory multi-Higgs SUSY model.
In a generic multi-Higgs model hard (CKM-like) $CP$ violation seems to
imply too large imaginary FCNCs mediated by the extra Higgs fields.
This fact strongly suggests a soft origin of $CP$ violation. Then the
problem of containing simultaneously FCNC and too large $CP$ violation
forces these models to have, for example, unobservable $CP$ asymmetries
in $B$ decays, a prediction that will be tested in the near future.

\vskip 1.0cm

{\bf Acknowledgments}

\vskip 0.5cm

We thank D. Chang, L. Hall, R. Mohapatra and A. Pomarol for helpful
suggestions and comments. The work of M. M. has been partially supported by
a grant from the Junta de Andaluc\'\i a (Spain). The work of A. R. was
supported by the NSF grant No. PHY9421385.

\vskip 1.0cm

{\bf Appendix:   Solutions to minimum equations}

\vskip 0.5cm

The vacuum expectation value of the scalar potential is
\ba
< V > & = &{1 \over 2} m_1^2 v_1^2 +
{1 \over 2} m_2^2 v_2^2 +
{1 \over 2} m_3^2 v_3^2 +
{1 \over 2} m_4^2 v_4^2 +
m_{12}^2 v_1 v_2 \cos \delta_2 +
 \nonumber \\
& + &  \epsilon_{PQ} m_{14}^2 v_1 v_4 \cos (\delta_4 + \alpha)
+ \epsilon_{PQ} m_{32}^2 v_3 v_2 \cos(\delta_3+\delta_2)
+ m_{34}^2 v_3 v_4 \cos(\delta_3+\delta_4)
+  \nonumber \\
& + &
{1 \over 32} (g^2 + g'^2) [ v_1^2 + v_3^2 - v_2^2 - v_4^2 ]^2 .
\label{eq:vevpot1}
\ea
The conditions at the minimum are
\ba
v_1 \frac {\partial V} {\partial v_1} & = &
    m_1^2 v^2_1 + m_{12}^2 v_1 v_2 \cos \delta_2
    + \epsilon_{PQ} m_{14}^2 v_1 v_4 \cos (\delta_4+\alpha)
    + v^2_1 g({\bf v}) = 0 \, , \nonumber\\
v_2 \frac {\partial V} {\partial v_2} & = &
    m_2^2 v^2_2 + m_{12}^2 v_1 v_2 \cos \delta_2
    + \epsilon_{PQ} m_{32}^2 v_3 v_2 \cos (\delta_3 + \delta_2)
    - v^2_2 g({\bf v}) = 0 \, , \nonumber\\
v_3 \frac {\partial V} {\partial v_3} & = &
    m_3^2 v^2_3 + \epsilon_{PQ} m_{32}^2 v_3 v_2 \cos (\delta_3 + \delta_2)
    + m_{34}^2 v_3 v_4 \cos (\delta_3 + \delta_4)
    + v^2_3 g({\bf v}) = 0 \, , \nonumber\\
v_4 \frac {\partial V} {\partial v_4} & = &
    m_4^2 v^2_4
    + m_{34}^2 v_3 v_4 \cos (\delta_3 + \delta_4)
    + \epsilon_{PQ} m_{14}^2 v_1 v_4 \cos (\delta_4 + \alpha)
    - v^2_4 g({\bf v}) = 0
\label{eq:minvevspq}
\ea
where $g({\bf v}) =   {1 \over 8} (g^2 + g'^2)
            [ v_1^2 + v_3^2 - v_2^2 - v_4^2 ]$,
and
\ba
- \frac {\partial V} {\partial \delta_2} & = &
    m_{12}^2 v_1 v_2 \sin \delta_2
    + \epsilon_{PQ} m_{32}^2 v_3 v_2 \sin (\delta_3 + \delta_2)
    = 0 \, , \nonumber\\
- \frac {\partial V} {\partial \delta_3} & = &
     \epsilon_{PQ} m_{32}^2 v_3 v_2 \sin (\delta_3 + \delta_2)
    + m_{34}^2 v_3 v_4 \sin (\delta_3 + \delta_4) = 0 \, , \nonumber\\
- \frac {\partial V} {\partial \delta_4} & = &
     m_{34}^2 v_3 v_4 \sin (\delta_3 + \delta_4)
    + \epsilon_{PQ} m_{14}^2 v_1 v_4 \sin (\delta_4 + \alpha)= 0 .
\label{eq:minphasespq}
\ea

First, we can estimate the relative sizes of VEVs. The solutions
to (\ref{eq:minvevspq}) are consistent with either one of the pairs of VEVs
$(v_1,v_2)$ or $(v_3,v_4)$ being suppressed by order $\epsilon_{PQ}$
with respect to the weak scale. Depending on the sizes of the unsupressed
parameters $m^2_1,m^2_3,m^2_{13}$
and $m^2_2,m^2_4,m^2_{24}$ the absolute minimum will prefer one of the above
choices. This can easily be seen from the following.
Imagine for a moment that $\epsilon_{PQ}=0$.
The equations in (\ref{eq:minvevspq}) reduce to two pairs of
equations, with first pair depending on the
ratio ${v_1 \over v_2}$ and $g({\bf v})$, and the second on
${v_3 \over v_4}$ and the same function $g({\bf v})$. Thus,
one pair of VEVs is forced to be zero.
Turning back on a small $\epsilon_{PQ}$, the terms in the Lagrangian
suppressed by $\epsilon_{PQ}$ can make the previously trivial
pair nonzero, but suppressed. The only question left is which pair of the
VEVs is small, and this will depend on the choice of the unsuppressed mass
parameters in the potential. We will assume that the parameters are such that
$(v_1,v_2)$ are unsuppressed (and of the order weak scale), while $(v_3,v_4)$
are of the order $\epsilon_{PQ}$ times
the weak scale. This assumption does not involve fine tuning
but only ``halves" the available parameter space.
For the phases $\delta_i$, it was shown in \cite{masi95} that in
the limit $\alpha=0$ there is no nontrivial solution
({\it i.e.}, $\delta_i$ different from 0 or $\pi$). It is easy to
see, however, that for $\alpha\neq 0$ the equations
(\ref{eq:minvevspq}) do {\it not} have this trivial solution, and
complex phases are guaranteed. In particular, for $\alpha=O(1)$ from
the two first equations in (\ref{eq:minphasespq}) it follows that
$\delta_2$ is of order $\epsilon_{PQ}^2$ (modulo $\pi$, depending on
the sign of $m^2_{12}$) and $\delta_3$ and $\delta_4$ are unsuppressed.

In summary, the structure of values of VEVs and their phases is
\ba
v_1 \, , v_2 & = & O(v) \nonumber \\
v_3 \, , v_4 & = & O(\epsilon_{PQ} v) \nonumber \\
\delta_2 & = & O(\epsilon^2_{PQ}) \nonumber \\
\delta_3 \, , \delta_4 & = & O(1)
\label{eq:structure1}
\ea
where $v$ denotes the weak scale.

We now explore this structure in more detail (at first order in
$\epsilon_{PQ}$). The first equation in (\ref{eq:minphasespq}) gives
$\delta_2$ to be of order $\epsilon^2_{PQ}$ up to\footnote{ In the
following we choose $m^2_{12}$ positive without loss of generality
and thus $\delta_2 \sim \pi + O(\epsilon^2_{PQ})$.}
a factor of $\pi$. Such $\delta_2$ does not contribute
to leading order (its contributions are O($\epsilon^2_{PQ}$))
to the minimum of the scalar potential (\ref{eq:vevpot1}),
and can be neglected in the rest of equations.
The minimum equations (\ref{eq:minvevspq}) to leading order are
\ba
v_1 \frac {\partial V} {\partial v_1} & = &
    m_1^2 v^2_1 + m_{12}^2 v_1 v_2
    + v^2_1 g_o({\bf v}) = 0 \, , \nonumber\\
v_2 \frac {\partial V} {\partial v_2} & = &
    m_2^2 v^2_2 + m_{12}^2 v_1 v_2
    - v^2_2 g_o({\bf v}) = 0 \, , \nonumber\\
v_3 \frac {\partial V} {\partial v_3} & = &
    m_3^2 v^2_3 - \epsilon_{PQ} m_{32}^2 v_3 v_2 \cos \delta_3
    + m_{34}^2 v_3 v_4 \cos (\delta_3 + \delta_4)
    + v^2_3 g_o({\bf v}) = 0 \, , \nonumber\\
v_4 \frac {\partial V} {\partial v_4} & = &
    m_4^2 v^2_4
    + m_{34}^2 v_3 v_4 \cos (\delta_3 + \delta_4)
    + \epsilon_{PQ} m_{14}^2 v_1 v_4 \cos (\delta_4 + \alpha)
    - v^2_4 g_o({\bf v}) = 0\;,
\label{eq:minvevspqsm}
\ea
where $g_o({\bf v}) =   {1 \over 8} (g^2 + g'^2)
            [ v_1^2 - v_2^2 ]$, and
\ba
- \frac {\partial V} {\partial \delta_3} & = &
     - \epsilon_{PQ} m_{32}^2 v_3 v_2 \sin \delta_3
    + m_{34}^2 v_3 v_4 \sin (\delta_3 + \delta_4) = 0 \, , \nonumber\\
- \frac {\partial V} {\partial \delta_4} & = &
     m_{34}^2 v_3 v_4 \sin (\delta_3 + \delta_4)
    + \epsilon_{PQ} m_{14}^2 v_1 v_4 \sin (\delta_4 + \alpha)= 0 .
\label{eq:minphasespqsm}
\ea
The first two equations in (\ref{eq:minvevspqsm})
are just the equations of the MSSM, and they fix $v_1$ and $v_2$ in the
usual way. Thus, we expect both $v_1$ and $v_2$ to be of the order
of weak scale and $\tan\beta=O(1)$ ({\it i.e.} no supression by
$\epsilon_{PQ}$, and no fine tuning producing large $\tan \beta$).

The goal now is to find the phases $\delta_3$ and $\delta_4$
in terms of quantities $c = \epsilon_{PQ} m^2_{32} v_3 v_2$,
$f= m^2_{34} v_3 v_4$ and $y = - \epsilon_{PQ} m^2 v_1 v_4$
and the angle $\alpha$ in order to eliminate them in the third
and fourth equation of VEVs in (\ref{eq:minvevspqsm}). We note
that the three quantities $c$, $f$, and $y$ are of order
$O(\epsilon^2_{PQ})$, and so we expect $\delta_3$ and $\delta_4$
unsupressed. In order to find $\delta_3$ and $\delta_4$ we
use a geometrical interpretation similar to the one devised in
\cite{masi95}. It is possible to see that the two equations
(\ref{eq:minphasespqsm}) define one of the objects shown
in Figure \ref{fig:trinontriv} (which one it is will depend whether
$1/c$, $1/$f and $1/y$ can form a triangle or not). The quantities
$p$ and $q$ there are not independent, and can be expressed in terms of
$c$, $f$, $y$ and $\alpha$. The difference between the two types
of solutions can be understood in the limit $\alpha \to 0$,
where only the trivial solutions $\delta_3 = 0$ and $\delta_4=\pi$
exist. For $\alpha=0$ the object in
Fig.~\ref{fig:trinontriv}(b) implies nonzero $\delta_3$ and
$\delta_4$, a type of solution which requires fine tuning between mass
parameters once it is substituted in the equations for $v_3$ and $v_4$.
In consequence, for small values of $\alpha$ only the solutions
of the type in Fig.~\ref{fig:trinontriv}(a) appear. When $\alpha$ is
nonzero the fine tuning is lifted, and both types of objects define
possible solutions to the minimum equations.

We performed numerical solutions to the above equations when
$\alpha \neq 0$ and large (order 1) and we found that the minima
satisfy the structure given in (\ref{eq:structure1}). For simplicity and
to illustrate the discussion above we will give the
equations for $v_3$ and $v_4$ at first order in $\alpha$.

{\bf Fig.~\ref{fig:trinontriv}(a):}
When $\alpha \to 0$ we see that $\delta_3 \to 0$ and
$\delta_4 \to \pi$, while
$1/p \to 1/c + 1/f$ and $1/q \to 1/y - 1/f$.
{}From the figure we first find cosines of the relevant angles
($\delta_3+\delta_4$, $\delta_3$ and $\delta_4+\alpha$) to leading
order in $\alpha$. Then we are in the position to find $v_3$ and $v_4$ by
substituting these expressions into the two last equations of
(\ref{eq:minvevspqsm})
\ba
v_3 \frac {\partial V} {\partial v_3} & = &
    m_3^2 v^2_3 - c [1 - {\alpha^2 \over 2}
({{ 1 \over {c}} \over
                 {{1 \over f} + {1 \over c} - {1 \over y}}})^2]
    - f [1 - {\alpha^2 \over 2}
({{ 1 \over {f}} \over
                 {{1 \over f} + {1 \over c} - {1 \over y}}})^2]
    + v^2_3 g({\bf v}) = 0 \, , \nonumber\\
v_4 \frac {\partial V} {\partial v_4} & = &
    m_4^2 v^2_4
    - f [1 - {\alpha^2 \over 2}
({{ 1 \over f} \over
                 {{1 \over f} + {1 \over c} - {1 \over y}}})^2]
    + y [1 - {\alpha^2 \over 2}
({ { 1 \over y} \over
                 {{1 \over f} + {1 \over c} - {1 \over y}}})^2]
    - v^2_4 g({\bf v}) = 0
\ea
where, again, $c = \epsilon_{PQ} m^2_{32} v_3 v_2$,
$f= m^2_{34} v_3 v_4$ and $y = - \epsilon_{PQ} m^2 v_1 v_4$.
These equations, although still complicated, can be solved
in $v_3$ and $v_4$, (remember that $v_1$ and $v_2$ are already fixed).
Then we can go back and find $\delta_3$ and $\delta_4$, thus completing
the search for the first case.

{\bf Fig.~\ref{fig:trinontriv}(b):}
In this case $1/p \to 1/y$ and $1/q \to 1/c$, while $\delta_3$
and $\delta_4$ tend to go to angles
in the triangle with sides $1/c$, $1/f$ and $1/y$ (we denote
this (order O(1)) asymptotic angles as $\delta^o_3$ and $\delta^o_4$).
We can again find the relevant angles to leading order in $\alpha$
and then find $v_3$ and $v_4$ by substituting these expressions into
the two last equations of (\ref{eq:minvevspqsm})
\ba
v_3 \frac {\partial V} {\partial v_3} & = &
    v^2_3 [ m^2_3 + \frac {m^2_{32} m^2_{34}} {m^2_{14}} {v_2 \over v_1} +
    g^0({\bf v}) ] +
        2 \alpha f \frac {\sin(\delta^o_3+\delta^o_4)}
                         {\tan \delta^o_3 \tan\delta^o_4}
        = 0 \, , \nonumber\\
v_4 \frac {\partial V} {\partial v_4} & = &
    v^2_4 [ m^2_4 + \frac {m^2_{34} m^2_{14}} {m^2_{32}} {v_1 \over v_2} -
    g^0({\bf v}) ] +
        2 \alpha f \frac {\sin(\delta^o_3+\delta^o_4)}
                         {\tan \delta^o_3 \tan\delta^o_4}
        = 0 \, , \nonumber\\
\ea
Remembering that $v_1$ and $v_2$ are already fixed in terms of
$m_1^2$, $m^2_2$ and $m^2_{12}$, we see clearly the fine tuning
for vanishing $\alpha$\cite{masi95}: the terms in square brackets would be
forced to vanish, implying two relations between mass parameters. However,
once we include $\alpha \neq 0$ these degeneracies get lifted.


\newpage
\textwidth 5.75in
\unitlength=1.00mm
\thicklines
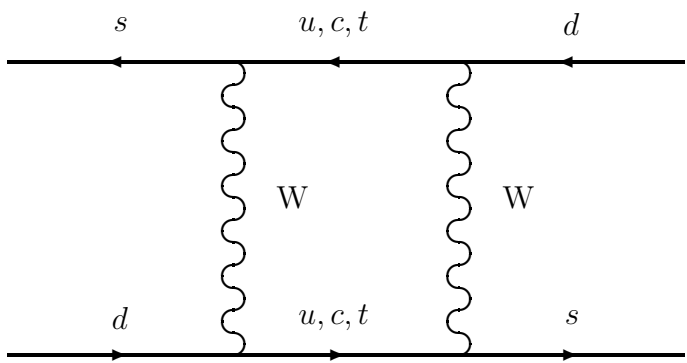
\begin{figure}
\begin{picture}(33.66,182.34)
\put(30.00,100.00){\line(1,0){90.00}}
\put(30.00,139.00){\line(1,0){90.00}}
\put(36.00,100.00){\vector(1,0){10}}
\put(65.00,100.00){\vector(1,0){10}}
\put(96.00,100.00){\vector(1,0){10}}
\put(53.00,139.00){\vector(-1,0){10}}
\put(82.00,139.00){\vector(-1,0){10}}
\put(113.00,139.00){\vector(-1,0){10}}
\put(46.00,105.00){\makebox(0,0)[rc]{$d$}}
\put(78.00,105.00){\makebox(0,0)[rc]{$u,c,t$}}
\put(106.00,105.00){\makebox(0,0)[rc]{$s$}}
\put(46.00,144.00){\makebox(0,0)[rc]{$s$}}
\put(78.00,144.00){\makebox(0,0)[rc]{$u,c,t$}}
\put(106.00,144.00){\makebox(0,0)[rc]{$d$}}
\put(60,101.5){\oval(3,3)[r]}
\put(60,104.5){\oval(3,3)[l]}
\put(60,107.5){\oval(3,3)[r]}
\put(60,110.5){\oval(3,3)[l]}
\put(60,113.5){\oval(3,3)[r]}
\put(60,116.5){\oval(3,3)[l]}
\put(60,119.5){\oval(3,3)[r]}
\put(60,122.5){\oval(3,3)[l]}
\put(60,125.5){\oval(3,3)[r]}
\put(60,128.5){\oval(3,3)[l]}
\put(60,131.5){\oval(3,3)[r]}
\put(60,134.5){\oval(3,3)[l]}
\put(60,137.5){\oval(3,3)[r]}
\put(90,101.5){\oval(3,3)[r]}
\put(90,104.5){\oval(3,3)[l]}
\put(90,107.5){\oval(3,3)[r]}
\put(90,110.5){\oval(3,3)[l]}
\put(90,113.5){\oval(3,3)[r]}
\put(90,116.5){\oval(3,3)[l]}
\put(90,119.5){\oval(3,3)[r]}
\put(90,122.5){\oval(3,3)[l]}
\put(90,125.5){\oval(3,3)[r]}
\put(90,128.5){\oval(3,3)[l]}
\put(90,131.5){\oval(3,3)[r]}
\put(90,134.5){\oval(3,3)[l]}
\put(90,137.5){\oval(3,3)[r]}
\put(70.00,121.00){\makebox(0,0)[rc]{W}}
\put(100.00,121.00){\makebox(0,0)[rc]{W}}
\end{picture}
\vspace{-8cm}
\caption{ Leading contribution to Re $\Delta M_{12}$.}
\label{fig:wexchange}
\end{figure}


\newpage
\textwidth 5.75in
\unitlength=1.00mm
\thicklines
\begin{figure}
\begin{picture}(33.66,182.34)
\put(30.00,100.00){\line(1,1){20.00}}
\put(30.00,140.00){\line(1,-1){20.00}}
\put(100.00,120.00){\line(1,1){20.00}}
\put(100.00,120.00){\line(1,-1){20.00}}
\multiput(50.00,120.00)(5,0){10}{\line(1,0){4.00}}
\put(45.00,105.00){\makebox(0,0)[rc]{$d$}}
\put(47.00,135.00){\makebox(0,0)[rc]{$s^c$}}
\put(105.00,105.00){\makebox(0,0)[rc]{$s$}}
\put(107.00,135.00){\makebox(0,0)[rc]{$d^c$}}
\put(78.00,125.00){\makebox(0,0)[rc]{$H$}}
\put(80.00,120.00){\vector(-1,0){5}}
\put(30.00,100.00){\vector(1,1){10}}
\put(30.00,140.00){\vector(1,-1){10}}
\put(100.00,120.00){\vector(1,1){10}}
\put(100.00,120.00){\vector(1,-1){10}}

\end{picture}
\vspace{-8cm}
\caption{ Leading contribution to Im $\Delta M_{12}$.}
\label{fig:treelevelfc}
\end{figure}
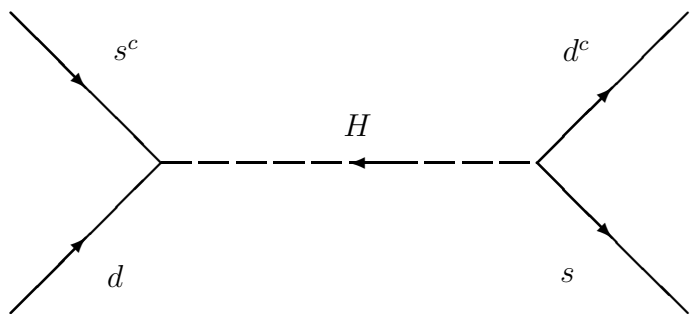


\newpage
\textwidth 5.75in
\unitlength=1.00mm
\thicklines
\begin{figure}
\begin{picture}(33.66,182.34)
\put(30.00,120.00){\line(1,0){100.00}}
\multiput(60.00,120.00)(5,5){4}{\line(1,1){4.00}}
\multiput(100.00,120.00)(-5,5){4}{\line(-1,1){4.00}}
\put(40.00,123.00){\makebox(0,0)[rc]{$s$}}
\put(120.00,123.00){\makebox(0,0)[rc]{$d$}}
\put(74.00,124.00){\makebox(0,0)[rc]{$\tilde{W}$}}
\put(90.00,124.00){\makebox(0,0)[rc]{$\tilde{H}$}}
\put(85.00,160.00){\makebox(0,0)[rc]{$g$}}
\put(72.00,135.00){\makebox(0,0)[rc]{$<H>$}}
\put(64.00,128.00){\makebox(0,0)[rc]{$\tilde{t}$}}
\put(97.00,134.00){\makebox(0,0)[rc]{$\tilde{t}^c$}}
\put(35.00,120.00){\vector(1,0){10}}
\put(105.00,120.00){\vector(1,0){10}}
\put(80.00,120.00){\vector(-1,0){10}}
\put(96.00,120.00){\vector(-1,0){10}}
\put(81.80,120.00){\makebox(0,0)[rc]{$\times$}}
\put(75.50,134.00){\makebox(0,0)[rc]{$+$}}
\put(86.80,115.00){\makebox(0,0)[rc]{$<H>$}}
\put(65,125){\vector(1,1){5}}
\put(95,125){\vector(-1,1){5}}

\put(80,141.5){\oval(3,3)[r]}
\put(80,144.5){\oval(3,3)[l]}
\put(80,147.5){\oval(3,3)[r]}
\put(80,150.5){\oval(3,3)[l]}
\put(80,153.5){\oval(3,3)[r]}
\put(80,156.5){\oval(3,3)[l]}
\put(80,159.5){\oval(3,3)[r]}
\end{picture}
\vspace{-8cm}
\caption{ Chargino contribution to $\epsilon'_K$.}
\label{fig:charginoep}
\end{figure}
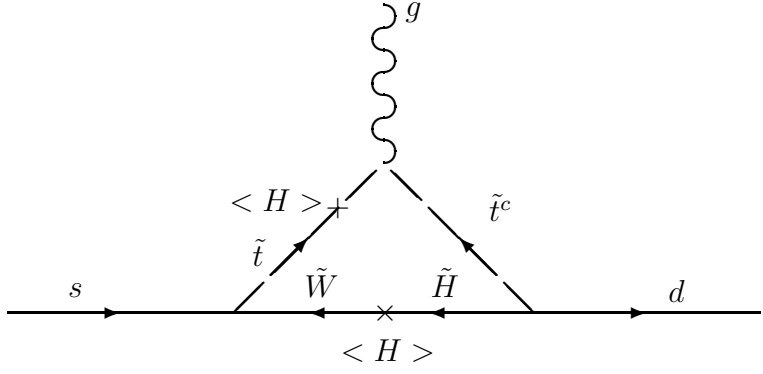


\newpage
\textwidth 5.75in
\unitlength=1.00mm
\thicklines
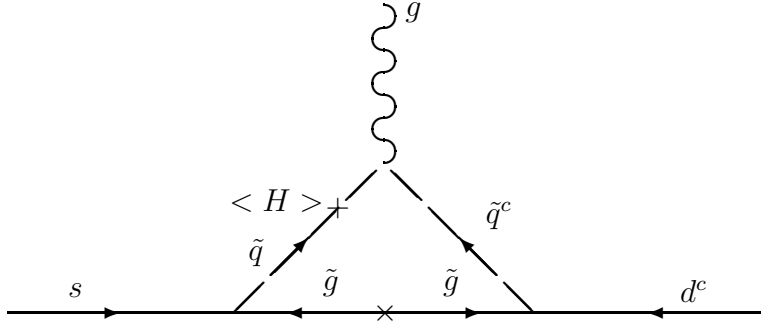
\begin{figure}
\begin{picture}(33.66,182.34)
\put(30.00,120.00){\line(1,0){100.00}}
\multiput(60.00,120.00)(5,5){4}{\line(1,1){4.00}}
\multiput(100.00,120.00)(-5,5){4}{\line(-1,1){4.00}}
\put(40.00,123.00){\makebox(0,0)[rc]{$s$}}
\put(123.00,123.00){\makebox(0,0)[rc]{$d^c$}}
\put(74.00,124.00){\makebox(0,0)[rc]{$\tilde{g}$}}
\put(90.00,124.00){\makebox(0,0)[rc]{$\tilde{g}$}}
\put(85.00,160.00){\makebox(0,0)[rc]{$g$}}
\put(72.00,135.00){\makebox(0,0)[rc]{$<H>$}}
\put(64.00,128.00){\makebox(0,0)[rc]{$\tilde{q}$}}
\put(97.00,133.00){\makebox(0,0)[rc]{$\tilde{q}^c$}}
\put(35.00,120.00){\vector(1,0){10}}
\put(125.00,120.00){\vector(-1,0){10}}
\put(77.00,120.00){\vector(-1,0){10}}
\put(83.00,120.00){\vector(1,0){10}}
\put(81.80,120.00){\makebox(0,0)[rc]{$\times$}}
\put(75.50,134.00){\makebox(0,0)[rc]{$+$}}
\put(80,141.5){\oval(3,3)[r]}
\put(80,144.5){\oval(3,3)[l]}
\put(80,147.5){\oval(3,3)[r]}
\put(80,150.5){\oval(3,3)[l]}
\put(80,153.5){\oval(3,3)[r]}
\put(80,156.5){\oval(3,3)[l]}
\put(80,159.5){\oval(3,3)[r]}
\put(65,125){\vector(1,1){5}}
\put(95,125){\vector(-1,1){5}}
\end{picture}
\vspace{-8cm}
\caption{ Gluino contribution to $\epsilon'_K$.}
\label{fig:gluinoep}
\end{figure}


\newpage
\textwidth 5.75in
\unitlength=1.00mm
\thicklines
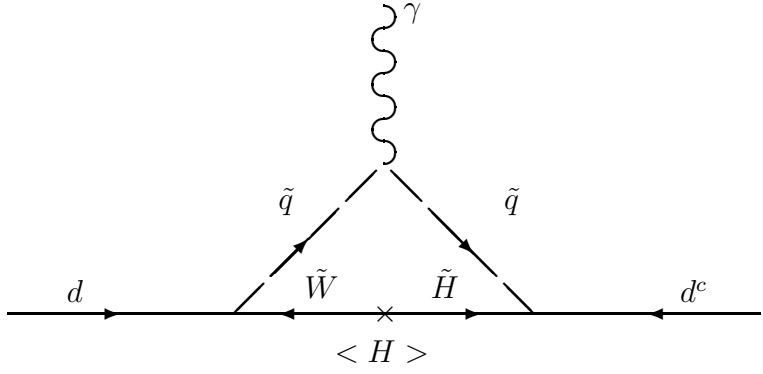
\begin{figure}
\begin{picture}(33.66,182.34)
\put(30.00,120.00){\line(1,0){100.00}}
\multiput(60.00,120.00)(5,5){4}{\line(1,1){4.00}}
\multiput(100.00,120.00)(-5,5){4}{\line(-1,1){4.00}}
\put(40.00,123.00){\makebox(0,0)[rc]{$d$}}
\put(123.00,123.00){\makebox(0,0)[rc]{$d^c$}}
\put(74.00,124.00){\makebox(0,0)[rc]{$\tilde{W}$}}
\put(90.00,124.00){\makebox(0,0)[rc]{$\tilde{H}$}}
\put(85.00,160.00){\makebox(0,0)[rc]{$\gamma$}}
\put(86.00,115.00){\makebox(0,0)[rc]{$<H>$}}
\put(68.00,135.00){\makebox(0,0)[rc]{$\tilde{q}$}}
\put(98.00,135.00){\makebox(0,0)[rc]{$\tilde{q}$}}
\put(81.80,120.00){\makebox(0,0)[rc]{$\times$}}
\put(80,141.5){\oval(3,3)[r]}
\put(80,144.5){\oval(3,3)[l]}
\put(80,147.5){\oval(3,3)[r]}
\put(80,150.5){\oval(3,3)[l]}
\put(80,153.5){\oval(3,3)[r]}
\put(80,156.5){\oval(3,3)[l]}
\put(80,159.5){\oval(3,3)[r]}
\put(65,125){\vector(1,1){5}}
\put(87,133){\vector(1,-1){5}}
\put(35.00,120.00){\vector(1,0){10}}
\put(125.00,120.00){\vector(-1,0){10}}
\put(76.00,120.00){\vector(-1,0){10}}
\put(83.00,120.00){\vector(1,0){10}}
\end{picture}
\vspace{-8cm}
\caption{ Chargino contribution to the neutron EDM.}
\label{fig:charginonedm}
\end{figure}


\newpage
\textwidth 5.75in
\unitlength=1.00mm
\thicklines
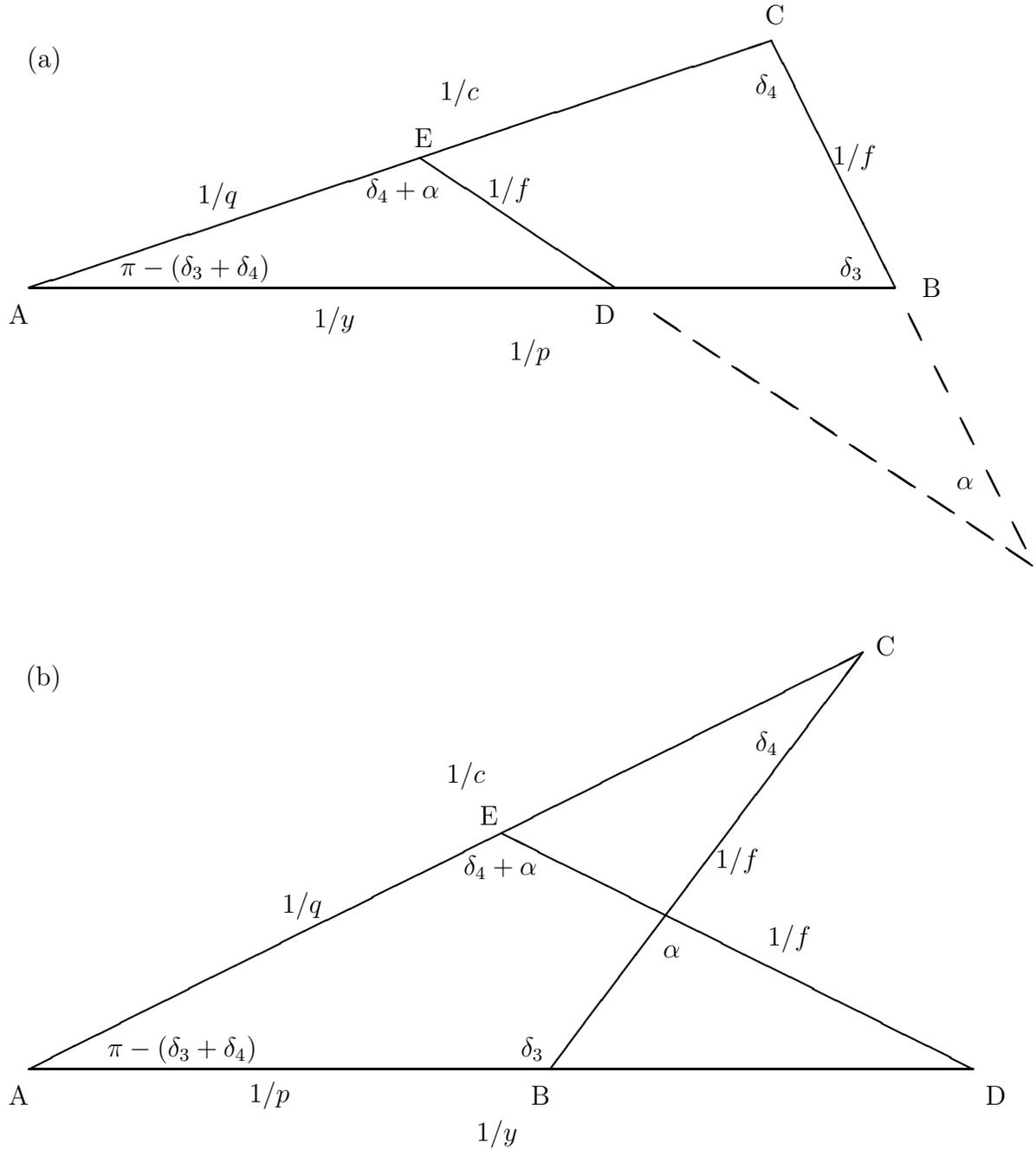
\begin{figure}
\begin{picture}(33.66,182.34)
\put(15.00,185.00){\makebox(0,0)[rc]{(a)}}
\put(10.00,150.00){\line(1,0){133.00}}
\put(10.00,150.00){\line(3,1){114.00}}
\put(70.00,170.00){\line(3,-2){30.00}}
\put(124.00,188.00){\line(1,-2){19.00}}
\multiput(106.00,146.00)(6,-4){10}{\line(3,-2){4.00}}
\multiput(145.00,146.00)(4,-8){5}{\line(1,-2){2.00}}
\put(138.00,153.00){\makebox(0,0)[rc]{$\delta_3$}}
\put(73.00,165.00){\makebox(0,0)[rc]{$\delta_4+\alpha$}}
\put(47.00,153.00){\makebox(0,0)[rc]{$\pi-(\delta_3+\delta_4)$}}
\put(125.00,181.00){\makebox(0,0)[rc]{$\delta_4$}}
\put(155.00,120.00){\makebox(0,0)[rc]{$\alpha$}}
\put(87.00,165.00){\makebox(0,0)[rc]{$1/f$}}
\put(140.00,170.00){\makebox(0,0)[rc]{$1/f$}}
\put(60.00,145.00){\makebox(0,0)[rc]{$1/y$}}
\put(90.00,140.00){\makebox(0,0)[rc]{$1/p$}}
\put(42.00,164.00){\makebox(0,0)[rc]{$1/q$}}
\put(79.00,180.00){\makebox(0,0)[rc]{$1/c$}}
\put(10.00,146.00){\makebox(0,0)[rc]{A}}
\put(100.00,146.00){\makebox(0,0)[rc]{D}}
\put(72.00,173.00){\makebox(0,0)[rc]{E}}
\put(150.00,150.00){\makebox(0,0)[rc]{B}}
\put(126.00,192.00){\makebox(0,0)[rc]{C}}
\put(10.00,30.00){\line(1,0){145.00}}
\put(90.00,30.00){\line(3,4){48.00}}
\put(10.00,30.00){\line(2,1){128.00}}
\put(155.00,30.00){\line(-2,1){72.50}}
\put(89.00,33.00){\makebox(0,0)[rc]{$\delta_3$}}
\put(125.00,80.00){\makebox(0,0)[rc]{$\delta_4$}}
\put(110.00,48.00){\makebox(0,0)[rc]{$\alpha$}}
\put(45.00,33.00){\makebox(0,0)[rc]{$\pi-(\delta_3+\delta_4)$}}
\put(88.00,61.00){\makebox(0,0)[rc]{$\delta_4+\alpha$}}
\put(15.00,90.00){\makebox(0,0)[rc]{(b)}}
\put(122.00,62.00){\makebox(0,0)[rc]{$1/f$}}
\put(130.00,50.00){\makebox(0,0)[rc]{$1/f$}}
\put(50.00,26.00){\makebox(0,0)[rc]{$1/p$}}
\put(85.00,20.00){\makebox(0,0)[rc]{$1/y$}}
\put(80.00,75.00){\makebox(0,0)[rc]{$1/c$}}
\put(55.00,55.00){\makebox(0,0)[rc]{$1/q$}}
\put(10.00,26.00){\makebox(0,0)[rc]{A}}
\put(90.00,26.00){\makebox(0,0)[rc]{B}}
\put(143.00,95.00){\makebox(0,0)[rc]{C}}
\put(160.00,26.00){\makebox(0,0)[rc]{D}}
\put(82.00,69.00){\makebox(0,0)[rc]{E}}
\end{picture}
\vspace{-1cm}
\caption{ The two possible geometrical objects which represent the
$CP$ nontrivial solution of equations (29) generated by
a nonzero soft phase $\alpha$. Each object
consists of two triangles, ABC and ADE.
The sides of the triangles are AB = $1/p$, BC = $1/f$, AC = $1/c$,
AD = $1/y$, DE = $1/f$, AE = $1/q$.
(a) The object is such that the sides $1/c$, $1/f$ and $1/y$
cannot form a triangle. (b) The object is such that the sides
$1/c$, $1/f$ and $1/y$ can form a triangle. }
\label{fig:trinontriv}
\end{figure}

\end{document}